\documentclass[seceq,supplement]{ptptex}

\usepackage{graphicx}
\usepackage{wrapft}

\title{
Disappearance of a Stacking Fault in Hard-Sphere Crystals under Gravity%
}

\author{
Atsushu \textsc{Mori},\footnote{E-mail: mori@opt.tokushima-u.ac.jp}%
Yoshihisa \textsc{Suzuki},
and
Shigeki \textsc{Matsuo}
}

\inst{
Institute of Technology and Science, The University of Tokushima,
Tokushima 770-8506, Japan
}

\abst{
In the first part of this paper, a review is given on the mechanism
for the disappearance of an intrinsic stacking fault in a hard-sphere
(HS) crystal under gravity, which we recently discovered by Monte Carlo
(MC) simulations
[A.~Mori \textit{et~al.}, \JL{J.~Chem.~Phys.,124,2006,17450};
\JL{Mol.~Phys.,105,2007,1377}].
We have observed, in the case of fcc (001) stacking,
that the intrinsic stacking fault running along an oblique direction
shrunk through the gliding of a Shockley partial dislocation at the lower
end of the stacking fault.
In order to address the shortcomings and approximations
of previous simulations,
such as the use of periodic boundary condition (PBC) and the fact that the fcc (001)
stacking had been realized by the stress from the small PBC box,
we present an elastic strain energy calculation for an infinite system
and a MC simulation result for HSs in a pyramidal pit under gravity.
In particular, the geometry of the latter has already been tested experimentally
[S.~Matsuo \textit{et~al.}, \JL{Appl.~Phys.~Lett.,82,2003,4283}].
The advantage of using a pyramidal pit as a template
as well as the feasibility of the mechanism we describe is demonstrated.
}

\begin{document}

\maketitle

\section{\label{sec:intro}
Introduction}
In 1957, the existence of the crystalline phase in a
hard-sphere (HS) system was found by Monte Carlo (MC) and molecular dynamics
(MD) simulations;~\cite{Wood1957,Alder1957}
the crystalline phase transition in the HS system is sometimes referred to as
the Alder transition. 
It was surprising that a system comprised merely of hard-core
repulsion exhibited a phase transition.
An intuitive understanding of the Alder transition is given by decomposing of
the entropy into a contribution due to the configurational variation of the HSs'
centers and a contribution due to the vibrational degree of freedom
around the HSs' equilibrium positions.
While the configurational entropy dominates in a disordered fluid phase,
the vibrational entropy dominates in the crystalline phase.
In the HS system the phase behavior is governed by the density;
the system is in the fluid phase at a density lower than
$\phi_f$ and in the crystalline phase of a face-centered cubic (fcc)
structure at a density higher than $\phi_s$.
Here, the particle density is expressed in terms of the volume fraction of the
HSs $\phi \equiv \pi\sigma^3N/6V$,
where $\sigma$ is the HS diameter, $N$ is the number of particles,
and $V$ is the total system volume.
A crystal of $\phi_s$ can coexist with a fluid of $\phi_f$ when the volume fraction
of the total system lies between $\phi_f$ and $\phi_s$.
The Hoover and Ree~\cite{Hoover1968} determined $\phi_f =0.494$ and $\phi_s = 0.545$
by a MC method in 1968.
Those values have been revised in the last decade to $\phi_f =0.491$ and
$\phi_s = 0.542$ by a MD simulation study of the direct two-phase coexistence
(i.e., the crystal/fluid interface).~\cite{Davidchack1998}
We note that even the first MD simulation of the HS crystal/fluid interface
was successfully performed within the last decade.~\cite{Mori1995}

The present situation is a little different from that in the early years.
In the 1960s and -70's, the existence of colloidal crystals drew much interest
as an experimental realization of the Alder transition.
The effective HS picture was proposed for charged colloids, which interacts
through a screened Coulomb potential where the interparticle interaction
is thus well described by a repulsive Yukawa form.~\cite{Wadachi1972}
Today, the HS system is not just an effective model of the colloids;
poly(methylmethacrylate) (PMMA) particles with stabilizing polymers grafted
on the particle's surface~\cite{Antl1986} were developed as HS suspensions.
The PMMA particles were dispersed in a compounded hydrocarbon medium
so that the HS nature with regards to the crystal-melt phase transition
was exhibited.~\cite{Antl1986,Pusey1986,Paulin1990,Underwood1994,Phan1996}

It should also be noted that colloidal crystals can possibly be used as
materials for photonic crystals.~\cite{Sakoda2001}
To realize a photonic band, the defects in the colloidal crystal should
be reduced.
To this end, many techniques have been developed.
One of them is colloidal epitaxy, in which a patterned substrate
is used as a template to fix the stacking direction to be
[001] in the sedimentation of the colloidal particles.~\cite{Blaaderen1997}
Various patterns for the template have been examined, but in this work only
a single pyramidal pit~\cite{Matsuo2003} is considered.

In order to improve the quality of the colloidal crystal, the effect of
gravity on the crystallinity of the colloidal crystal must be taken into
account.
Zhu \textit{et~al.}~\cite{Zhu1997} performed a colloidal crystallization
on the Space Shuttle and concluded that under microgravity,
a random hexagonal close pack (rhcp) was found.
On the oher hand, under normal gravity, the colloidal crystal formed by
sedimentation is a rhcp/fcc mixture although the colloids sometimes freeze
into a glassy state.
Here, rhcp is the random staking of hexagonal planes (fcc \{111\} or hcp
(0001)); viewed along $\langle$111$\rangle$, the fcc is of ABCABC$\cdots$
type, and the hcp is of ABAB$\cdots$ type while the rhcp corresponds to
a random sequence.
In other words, the stacking disorder can be reduced by gravity.
We have, however, not found a final answer to the stacking sequence of
colloidal crystals under gravity.
Although the gravitational constant $g^* \equiv mg\sigma /k_BT$ was different,
Kegel and Dhont~\cite{Kegel2000} observed a faulted twinned fcc under gravity.
Here, $m$ is the particle's mass, $g$ is the acceleration due to gravity,
$\sigma$ is the HS diameter, and $k_BT$ is the temperature multiplied by
Boltzmann's constant.
We note that the temperature $T$ should be defined from the thermal motion
of the dispersing particles.

This paper focuses on the mechanism of the disappearance of the stacking
disorder in a HS crystal under gravity.
We have already demonstrated the disappearance of stacking disorder
in a HS crystal under gravity by MC simulations.~\cite{Mori2006JCP}
We have examined closely where the shrinking of an intrinsic
stacking fault was observed.~\cite{Mori2007}
In \S~\ref{sec:PBC}, we review those simulation results, point out their
shortcomings, and reiterate the mechanism of the defect disappearance
that had been found.
The objective of this paper is to provide new results to complement
these shortcomings.
The systems~\cite{Mori2006JCP,Mori2007} were highly stressed because of
their smallness.
We preset some results of the elastic energy calculation for an infinite
system in \S~\ref{sec:elastic}.
The periodic boundary condition (PBC) applied in the previous
simulations~\cite{Mori2006JCP,Mori2007} was one of the artifacts that needed
to be addressed.
In \S~\ref{sec:pyramidal} we present results of the MC simulation for
realizable boundary conditions by simulating HSs in a pyramidal pit.~\cite{Matsuo2003}

\section{\label{sec:PBC}
Monte carlo simulation under the PBC}
\begin{wrapfigure}[25]{r}{\halftext}
\begin{center}
\centerline{\includegraphics[width=6cm]{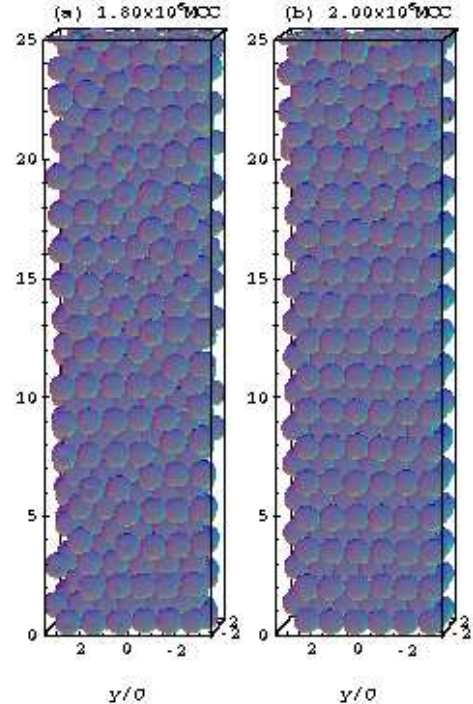}}
\caption{\label{fig:PBC}
Snapshots at (a) the beginning and (b) the end of a MC
simulation~\cite{Mori2006JCP} with $g^*$ fixed at 0.9.}
\end{center}
\end{wrapfigure}
In previous simulations,~\cite{Mori2006JCP} we presented results
for two system sizes.
One contained $N$=1664 particles and the other $N$=3744.
In the former, fcc (001) stacking occurred, while in the latter,
stacking of the hexagonal layers took place.
We concentrated on the former~\cite{Mori2007} because the defects' disappearance
was observed in this case.
Although the side length equaled $4a_0$, six particles lay along a side.
Here, $a_0=1.57\sigma=(4\sigma^3/\rho_s)^{1/3}$
was the fcc lattice constant at the crystal/fluid coexistence of Hoover and
Ree's value~\cite{Hoover1968} with $\rho$ denoting the particle density
[$\phi \equiv (\pi/6)\rho \sigma^3$].
Six unit cells lay along the diagonal direction.
In other words, the [010] and [100] are directed
to the diagonal directions.~\cite{Mori2006STAM}
This indicated that the crystal at the bottom was at a higher pressure than the
crystal-fluid coexistence pressure
[the Hoover and Ree's value is $(P\sigma^3/k_BT)_{coex}=11.75$~\cite{Hoover1968}
and Davidchack and Laird's value 11.55~\cite{Davidchack1998}]
in accordance with the mechanical balance equation
\begin{equation}
\label{eq:mech}
\partial P/\partial z=-mg\rho(z)
\end{equation}
at the sedimentation equilibrium, where $\rho(z)$ is the particle density
on a coarse-scale at altitude $z$.
Also, (\ref{eq:mech}) implies that the lattice spacing along $z$ direction
varies and that the unit cell is compressed along the vertical direction,
i.e., the crystal is no longer a cubic system, as was previously
confirmed.~\cite{Mori2006STAM}
The disappearance of the stacking fault occurred under such artificially
stressed condition.
Such stress, however, plays a central role in defect reduction.
As shall be described below, the (001) stacking, in which the shrinking of
the stacking fault occurred,~\cite{Mori2007} was thereby induced.
Moreover, some researchers made an efforts to realize the (001) growth, such as
the colloidal epitaxy,~\cite{Blaaderen1997} with the understanding that
the stacking sequence for this growth is unique. 
Concentrating on phenomena occurring under the (001) stacking without
particularly concerning ourselves with the experimental realization of this
stacking is one of paths to elucidation of the defect reduction mechanism.

In previous simulations,~\cite{Mori2006JCP,Mori2007} the gravitational constant,
$g^*$, was changed in a step-wise fashion in order to avoid trapping the system
in a metastable state such as a polycrystalline state.~\cite{Yanagiya2005}
We set $g^*$ to a certain value for 2$\times 10^5$ MC cycle (MCC), and then
changed $g^*$ by an amount of $\Delta g^*$ = 0.1, where one MCC is defined so as
to include, on average, one position move per particle.
Transformation of a defective crystal into a less-defective crystal was observed
for the case of (001) stacking for $g^*$ around $g^*=0.9.$~\cite{Mori2006JCP}
While $g^*$ was kept at 0.9 for the (001) stacking case, we found that the shrinking
of an intrinsic stacking fault was mediated by the glide of the Shockley partial
dislocation terminating at the bottom end of the stacking fault running in the
oblique direction.~\cite{Mori2007}
In Fig.~\ref{fig:PBC} we see a disappearance of the stacking faults,
which run from the lower-left to the middle-right, 
[see, Ref.~\citen{Mori2007} for details of the shrinking of the intrinsic
stacking fault].
We stress that the advantage of the (001) stacking is not only the uniqueness
of the stacking sequence but also the glide mechanism of the Shockley partial
dislocation.

\section{\label{sec:elastic}
Elastic strain energy consideration}
\begin{wrapfigure}[17]{r}{\halftext}
\begin{center}
\centerline{\includegraphics[width=6cm]{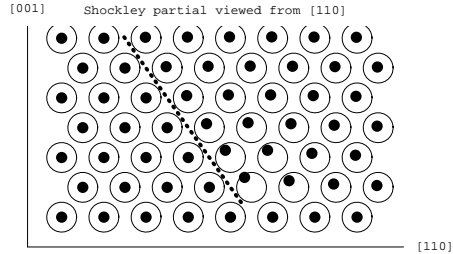}}
\caption{\label{fig:ISF}
Illustration of an intrinsic stacking fault.
Particles in the distorted crystal are indicated by dots, and the regular
lattice positions are indicated by open circles.
The dotted line indicates the stacking fault.
For simplicity, particles outside the portion right of stacking fault are
not displaced in this illustration.}
\end{center}
\end{wrapfigure}
In this section, we present the elastic energy calculation for a system
including an intrinsic stacking fault running along the [111].
The lower end of the stacking fault is terminated by a Shockley
partial dislocation, such as the one shown in Ref.~\citen{Mori2007}
and illustrated in Fig.~\ref{fig:ISF}.
In this paper, we incorporated the effect of gravity as buoyancy
due to the particle deficiency accompanied by the dislocation core.
The particle deficiency is one-third in Fig.~\ref{fig:ISF}.
Though the crystal was indeed strained due to gravity,~\cite{Mori2006STAM}
which coupled with the stress due to the Shockley partial
dislocation to help promote the shrinking of the intrinsic stacking
fault, we can understand the shrinking of the intrinsic stacking fault
without considering the cross coupling between these stresses.

\begin{wrapfigure}[12]{r}{\halftext}
\begin{center}
\centerline{\includegraphics[width=6cm]{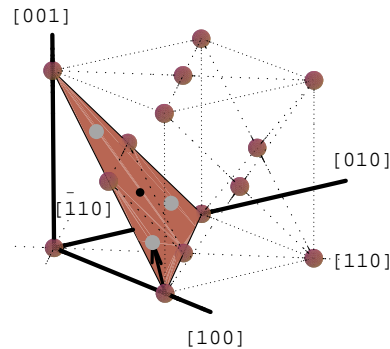}}
\caption{\label{fig:111}
The Burgers vector $\mbox{\boldmath $b$}^I = (1/6)[\bar{2}11]$ (arrow)
and the (111) plane (painted).
The arrow connects a lattice position to an adjacent lattice position,
say lattice point A to lattice point B.}
\end{center}
\end{wrapfigure}
The Shockley partial dislocation terminating an intrinsic stacking
fault running along the [111] (as shown in Fig.~\ref{fig:ISF}) is
defined by the Burgers vector $\mbox{\boldmath $b$}^I = (1/6)[\bar{2}11]$
$\equiv -\mbox{\boldmath $a$}_1/3 + \mbox{\boldmath $a$}_2/6
+ \mbox{\boldmath $a$}_3/6$ with $\mbox{\boldmath $a$}_1$,
$\mbox{\boldmath $a$}_2$, and $\mbox{\boldmath $a$}_3$ being the lattice vectors,
which is shown by the arrow in Fig.~\ref{fig:111}.
We can understand the partial dislocation by the decomposition of a perfect dislocation.
In Fig.~\ref{fig:111}, the Burgers vector of a perfect dislocation,
$\mbox{\boldmath $b$} =(1/2)[\bar{1}10]$, is decomposed as
$\mbox{\boldmath $b$} = \mbox{\boldmath $b$}^I + \mbox{\boldmath $b$}^{II}$
with $\mbox{\boldmath $b$}^{II} = (1/6)[\bar{1}2\bar{1}]$ as was done in
Ref.~\citen{Hirth}.

We calculated the elastic energy due to a dislocation running
along a unit vector $\mbox{\boldmath $\xi $}$ using the following
formula from isotropic, linear elastic theory~\cite{Hirth}
\begin{equation}
W(R) = \frac{\mu b^2}{4\pi}
\left(\cos^2 \theta + \frac{\sin^2 \theta}{1-\nu} \right)
\ln \left(\frac{\alpha R}{b}\right),
\label{eq:self-energy}
\end{equation}
where $\mu$, $\nu$, and $\theta$ are the shear modulus, the Poisson ratio,
and the angle between the Burgers vector $\mbox{\boldmath $b$}$ and
$\mbox{\boldmath $\xi $}$, respectively, and $b$ denotes $|\mbox{\boldmath $b$}|$.
Here, with $\alpha$, which is several tenths, the radius of defined core region
is $r_0 \equiv b/\alpha$, and $R$ is the dimension over which the elastic field
expands, which we can usually identify with the crystallite or grain radius.
In the present case, where the stacking fault is running upward along the
$\langle 111 \rangle$ starting at the position of the Shockley partial dislocation,
we simply set $R$ to the distance from the upper boundary to the Shockley
partial dislocation.
(Note that the dependence of the geometry of the boundary or the shape of
the crystallite is ignored.)
Substituting $\mbox{\boldmath $b$}^I$ for $\mbox{\boldmath $b$}$
(i.e., $|\mbox{\boldmath $b$}^I| = a/\sqrt{6}$ with $a$ being the fcc lattice constant)
and $\mbox{\boldmath $\xi$} = (1/\sqrt{2})[\bar{1}10]$ (i.e., $\theta = \pi/6$),
we obtain an elastic energy of
\begin{equation}
\label{eq:Uel}
U_{el} = \frac{\mu a^2}{96\pi} \left( 3+\frac{1}{1-\nu} \right)
\ln \left( \frac{\sqrt{6} \alpha R}{a} \right).
\end{equation}
Hereafter, we consider a system of unit length thickness perpendicular to
Fig.~\ref{fig:ISF}.
The core region is defined so that linear elastic theory is still valid outside
that region.
Borrowing the empirical result for metals,~\cite{Hirth}
the core energy $U_{core}$ is proportional to $\mu b^2$.
Let us write $U_{core} = \beta \mu b^2$ with $\beta$ being a certain constant
of order less than unity.
We have
\begin{equation}
\label{eq:Ucore}
U_{core} = \beta \mu a^2/6,
\end{equation}
($\mbox{\boldmath $b$} = \mbox{\boldmath $b$}^I$) for the intrinsic
stacking fault.
Note that (\ref{eq:Ucore}) is independent of $R$.

We should now consider the stacking fault energy.
This quantity, as well as the elastic and core energies (rigorously, the free
energies), is an entropic contribution resulting from the variation of the
vibrational mode distribution.
Though the core energy has not been calculated, the shear modulus $\mu$ was
calculated by a MC simulation~\cite{Frenkel1987} and density functional
theory,~\cite{Laird1992} and the stacking fault energy $\gamma_{sf}$ was
calculated by a MC simulation.~\cite{Pronk1999}
The quantity, $\mu \sigma^3/k_BT$ for the HS crystal ranges between 50 and 100,
depending on the particle number density, $\rho$.
This range of $\mu$ corresponds to $\rho \sigma^3 \cong 1.06 - 1.13$
($a/\sigma$ $\cong$ 1.55 -- 1.52) where the disappearance of the stacking
disorder was observed in the MC simulations.~\cite{Mori2006JCP,Mori2007}
The stacking fault interfacial energy per unit area, $\gamma_{sf} \sigma^2/k_BT$,
was (at most) of the order $10^{-4}$ at $\rho \sigma^3 = 1.10$.
The difference of the orders in $\mu$ and $\gamma_{sf}$ has a crucially important
meaning; if the perfect dislocation is decomposed into two partial dislocations
connected by a stacking fault, the separation between two partial dislocations
would be much longer than the order of the crystallite radius, which is typically
a few to several hundred lattice spacings.
The total stacking fault energy is obtained by multiplying the length of
the stacking fault, which is $\zeta R$; the proportional
constant $\zeta$ depends on the geometry of the boundary.
Thus, we have $U_{sf} = \zeta \gamma_{sf} R$.

The gravitational energy is solely given by $U_g(R) - U_g(0) = m \rho a^2 R/3$,
where we have neglected the dependence of the particle density on the altitude.
This approximate treatment was consistent with the treatment where we neglected
the deformation according to (\ref{eq:mech}).
Here $U_g(R)$ is the gravitational energy where the lower end of the stacking
fault is located from the upper boundary of the crystallite; so $U_g(0)$ vanished
because the stacking fault went out of the crystallite when $R=0$.
The sum of gravitational energy and the stacking fault energy was thus
a linear function
\begin{equation}
\label{eq:Usfg}
U_{sf}+U_g = (\zeta \gamma_{sf} + m \rho a^2/3) R.
\end{equation}

The total energy, $U_{el}+U_{core}+U_{sf}+U_g$, is comprised of a logarithmic term
(\ref{eq:Uel}), a constant (\ref{eq:Ucore}), and a linear term (\ref{eq:Usfg}).
The coefficient of the logarithmic term is positive and that
of the linear term is also apparently positive.
This means that as $R$ decreases, the energy decreases.
We have shown by an elastic calculation that in the (001) stacking, gravity provided
a driving force that promoted the glide of the Shockley partial dislocation upwards.

\section{\label{sec:pyramidal}
Monte carlo simulation of hard spheres in a pyramidal pit}
\begin{wrapfigure}[15]{r}{\halftext}
\begin{center}
\centerline{\includegraphics[width=5cm]{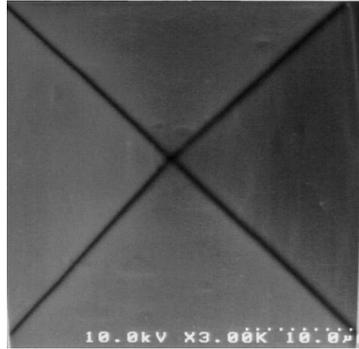}}
\caption{\label{fig:Si111}
A top view of a pyramidal pit made of four Si \{111\} faces
(scanning microscopy image).}
\end{center}
\end{wrapfigure}
In the preceding two sections, we considered two systems with artificial
boundary conditions.
In contrast, in this section, we treat a realistic boundary condition.
Mastsuo~\textit{et~al.}~\cite{Matsuo2003} used a pyramidal pit and groove
to fix the stacking sequence.
A pyramidal pit can be made by the anisotropic etching of the Si (001) surface
(Fig.~\ref{fig:Si111}).

\begin{figure}[h]
\begin{center}
\centerline{\includegraphics[width=12cm]{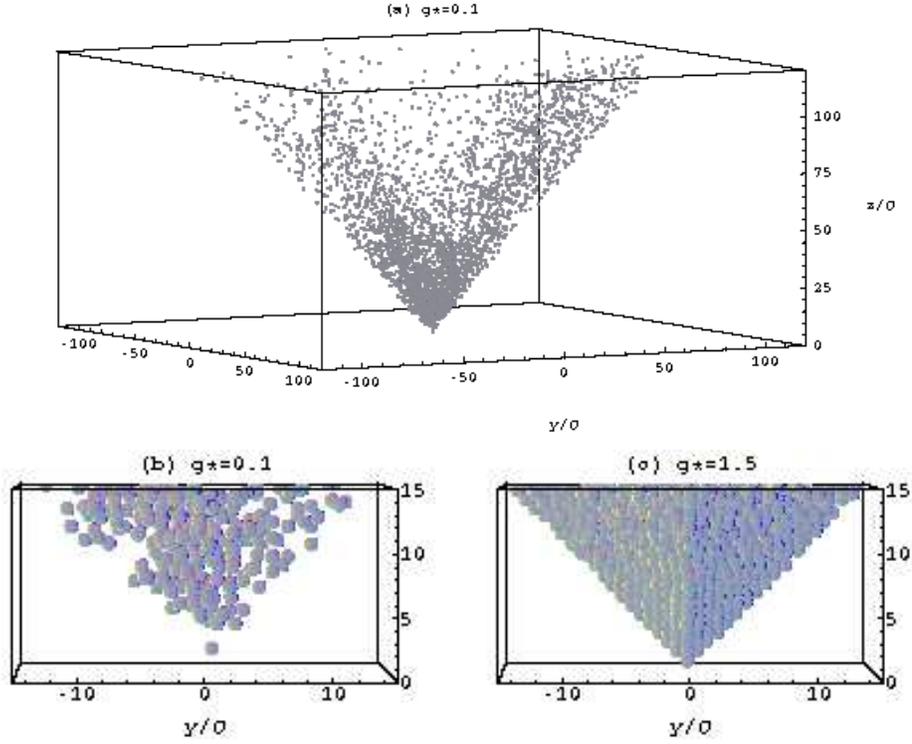}}
\caption{\label{fig:pit}
Snapshots of MC simulation of the HSs in a pyramidal pit.
(a) An overview at $g^*$=0.1, where edge wetting is seen,
(b) a magnification of (a), and (c) a magnified snapshot at $g^*$=1.5,
where crystalline order was observed.}
\end{center}
\end{figure}

Contrary to the simulations with PBC, the crystallinity was less sensitive
to how $g^*$ was controlled.
This may be due to wetting on the edges of the pyramidal pit
(Fig.~\ref{fig:pit}(a)).
In Fig.~\ref{fig:pit}, snapshots of MC simulations at $g^*=0.1$ and 1.5 are shown;
though $\Delta t$ = 2 and $4 \times 10^5$MCC, where $g^*$ was kept constant, were
also tested, only the results for $\Delta t$ = $1 \times 10^5$MCC are shown.
Figure~\ref{fig:pit}(b) shows that despite the edge wetting, the particles at the
bottom were not crystallized.
On the other hand, the particles were crystallized at the bottom of the pit, particularly,
in the fcc (001) stacking.
In the pyramidal pit made of the \{111\} faces, which can be experimentally
generated, the (001) stacking has been confirmed, as was already experimentally
demonstrated.~\cite{Matsuo2003}
Unfortunately, looking over the snapshots, we could not find the stacking faults, and
thus, the glide mechanism of the stacking fault disappearance was not observed.
In our opinion, however, this was an indication of this method's robustness 
against stacking disorder; the epitaxial growth at the late stage started from
the edges of the pit where the wetting occurred.
In addition, the geometry of the boundary was advantageous; even if a crystallite
mismatched its crystallographic orientation or lateral position with the formed
crystal as it nucleates at the bottom, the crystallite could slide along the wall.
In other words, the simulation results did not rule out the glide mechanism of
a stacking fault but instead emphazised its role.
%

\section{Discussions}
Here we compare two simulations.
Under PBC, the configuration is equivalent to the dislocation array of
the same Burgers vector.
On the other hand, owing to the concept of the mirror image, the pyramidal
pit system is equivalent to an array of opposite charges.
Whereas the former is less mobile, the dislocation in the latter more
easily disappears, which can be understood by pair annihilation of opposite
charges.
The glide mechanism is likely implicated because it was observed in the former
configuration.

The cross term between the gravity-induced elastic field and the stress due to
dislocation has not been treated in this paper.
From (\ref{eq:mech}), the particle density decreases as the altitude increases.
This means that the buoyancy of the dislocation core is promoted as the core
goes up because the higher the altitude is, the looser the packing is.
Essentially the same argument can be valid for the strain energy to due the dislocation.

\section{Concluding remarks}
We presented two simulation results and one theoretical calculation.
All of our results support the glide mechanism for the disappearance of the
stacking disorder in a hard-sphere crystal under gravity.
We wish to emphasize that although this mechanism is not the final unique answer,
it is likely to occur.

In a theoretical elastic calculation, we can understand the glide mechanism
without taking into account the cross term between the gravity-induced elastic
field and the stress due to the dislocation.
That is, $\partial \sigma_{ij}/\partial x_i = 0$ has been solved, where
$\sigma_{ij}$ is the stress tensor.
The rigorous treatment is to solve $\partial \sigma_{ij}/\partial x_i + f_i= 0$
with $f_i$ denoting the external force (gravitational force in the present case).
This calculation is currently underway using the results from Ref.~\citen{Mori2006STAM}.


%

\section*{Errata [Prog.~Theor.~Phys.~Suppl.\ \textbf{178} (2009), 33.]}
Both in Eq. (3.2) and on the line 5 on page 38 (\pageref{eq:Usfg}th page),
we should add a prefactor ($1/2\sqrt{2})$ on
$m \rho a^2/3$ and 
$m \rho a^2R/3$, respectively.
This factor arizes from the area per [$\bar{1}$10] line
on (110) plane, which is $a^2/2\sqrt{2}$.
We can also drive this by considering the
number of particle on [$\bar{1}$10] per unit length,
i.e, $U_g=m(\sqrt{2}/a)/3=(1/2\sqrt{2})m\rho a^2/3$
because $\rho=4/a^3$ for fcc lattice.

\begin{thebibliography}{99}
  
\bibitem{Wood1957} W.~W.~Wood and J.~D.~Jacobson, \JL{J.\ Chem.\ Phys.,27,1957,1207}.

\bibitem{Alder1957} B.~J.~Alder and T.~E.~Wainwright, \JL{J.\ Chem.\ Phys.,27,1957,1208}.

\bibitem{Hoover1968} W.~G.~Hoover and F.~H.~Ree, \JL{J.\ Chem.\ Phys.,49,1968,3609}.

\bibitem{Davidchack1998} R.~L.~Davidchack and B.~B.~Laird,
\JL{J.\ Chem.\ Phys.,108,1998,9452}.

\bibitem{Mori1995} A.~Mori, R.~Manabe and K.~Nishioka, \PRE{51,1995,R3831}.

\bibitem{Wadachi1972} M.~Wadati and M.~Toda, \JPSJ{32,1972,1147}.

\bibitem{Antl1986} L.~Antl, J.~W.~Goodwin, R.~D.~Hill, R.~H.~Ottweil,
S.~M.~Owens, S.~Parworth and J.~W.~Waters, \JL{Colloids\ Surf.,17,1986,67}.

\bibitem{Pusey1986} P.~N.~Pusey and W.~van~Megen, \JL{Nature,320,1986,340}.

\bibitem{Paulin1990} S.~E.~Paulin and B.~J.~Ackerson, \PRL{64,1990,2663}.

\bibitem{Underwood1994} S.~M.~Underwood, J.~R.~Taylor and W.~van~Megen,
\JL{Langmuir,10,1994,3550}.

\bibitem{Phan1996} S.~E.~Phan, W.~B.~Russel, Z.~Cheng, J.~Zhu, P.~M.~Chaikin,
J.~H.~Dunsmur and R.~H.~Ottewil, \PRE{54,1996,6633}.

\bibitem{Sakoda2001} K.~Sakoda, \textit{Optical Properties of Photonic Crystal}
(Springer-Verlag, Bargin, 2001).

\bibitem{Blaaderen1997} A.~van~Blaaderen, R.~Ruel and P.~Wiltzius,
\JL{Nature,385,1997,321}.

\bibitem{Matsuo2003} S.~Matsuo, T.~Fujine, K.~Fukuda, S.~Joudokazis and
H.~Misawa, \JL{Appl.\ Phys.\ Lett.,82,2003,4283}.

\bibitem{Zhu1997} J.~Zhu, M.~Li, R.~Rogers, W.~Mayer, R.~H.~Ottewil,
STS-73 Space Shuttle Crew, W.~B.~Russel and P.~M.~Chaikin,
\JL{Nature,387,1997,883}.

\bibitem{Kegel2000} W.~K.~Kegel and J.~K.~G.~Dhont,
\JL{J.\ Chem.\ Phys.,112,2000,3431}.

\bibitem{Mori2006JCP} A.~Mori, S.-i.~Yanagiya, Y.~Suzuki, T.~Sawada and
K.~Ito, \JL{J.\ Chem.\ Phys.,124,2006,174507}.

\bibitem{Mori2007} A.~Mori, Y.~Suzuki, S.-i.~Yanagiya, T.~Sawada
and K.~Ito, \JL{Mol.\ Phys.,105,2007,1377}.

\bibitem{Mori2006STAM} A.~Mori, S.-i.~Yanagiya, Y.~Suzuki, T.~Sawada and
K.~Ito, \JL{Sci.\ Technol.\ Avd.\ Mater.,7,2006,296}.

\bibitem{Yanagiya2005} S.-i.~Yanagiya, A.~Mori, Y.~Suzuki, Y.~Miyoshi,
M.~Kasuga, T.~Sawada, K.~Ito and T.~Inoue,
\JL{Jpn.\ J.\ Appl.\ Phys. (Part 1),44,2005,5113}.

\bibitem{Hirth} J.~P.~Hirth and J.~Lothe, \textit{Theory of dislocations}
(Krieger, Melabar, 1982).

\bibitem{Frenkel1987} D.~Frenkel and A.~J.~C.~Ladd,
\PRL{59,1987,1169}.

\bibitem{Laird1992} B.~B.~Laird,
\JL{J.\ Chem.\ Phys.,97,1992,2699}.

\bibitem{Pronk1999} S.~Pronk and D.~Frenkel,
\JL{J.\ Chem.\ Phys.,110,1999,4589}.


\end{thebibliography}
\end{document}